\documentstyle[prb,aps,floats,epsf]{revtex} 

\begin{document}

\title{Unitary Limit of Spin-Orbit Scattering in Two-Dimensional $s$-
and $d$-Wave Superconductors}
\author{Claudio Grimaldi} 
\address{\'Ecole Polytechnique F\'ed\'erale de Lausanne,
D\'epartement de Microtechnique IPM, CH-1015 Lausanne, Switzerland}

\date{\today} 
\maketitle 

\begin{abstract}
Non-magnetic impurities affect the paramagnetic response of superconductors
via the associated spin-orbit interaction which, when the non-magnetic impurity
is close to the unitary limit,
must be treated beyond the classical Born approximation. Here 
the Zeeman response of two-dimensional $s$- and $d$-wave superconductors is
calculated
within the self-consistent $T$-matrix formulation for both impurity
and spin-orbit scatterings. It is shown that at the unitary limit, for
which the spin-orbit scattering is maximum, the spin-up and spin-down 
channels becomes decoupled implying full Zeeman splitting of the
quasiparticle excitations. These results could be used to test the
unitary scattering hypothesis in high-$T_c$ superconductors.
\\
PACS number(s): 74.20.Fg, 71.70.Ej, 74.62.Dh
\end{abstract}

\section{Introduction}
\label{intro}
The response of a superconductor to defects and/or impurities
brings important informations on the nature of the superconducting state
and has been the subject of an enormous amount of theoretical
and experimental research.\cite{flatte} The effect of impurities on the superconducting
state depends crucially on the nature of the impurity (magnetic or 
non-magnetic) and on the symmetry of the order parameter. For isotropic
$s$-wave superconductors, the response to a non-magnetic impurity is 
feeble,\cite{anderson} while the effect of a magnetic one is 
dramatic.\cite{abri,shiba}
Instead when the order parameter is anisotropic, like in $d$-wave 
high-$T_c$ superconductor, the response to disorder is 
always dramatic,\cite{hirsch,sunmaki} leading to Kondo-like effects in case
of magnetic scattering potentials,\cite{polko}
or resonant behaviors for non-magnetic impurities close to 
the unitary limit.\cite{balatsky} 
For high-$T_c$ superconductors therefore it is experimentally more 
difficult to establish whether the impurity acts effectively as a magnetic or
a non-magnetic scattering potential. For example, recent scanning tunneling
microscope images of the local tunneling conductivity around a Zn
impurity in Bi2212,\cite{hudson} have been fitted by both 
non-magnetic\cite{haas} and
magnetic impurity models.\cite{polko} 

A topic which could be helpful to clarify the effective nature of disorder
in high-$T_c$ oxides is the analysis of the response to some external applied
perturbation.
The aim of this paper is to show some important consequences of having strong
non-magnetic impurities on the Zeeman response of a $s$- or
$d$-wave superconductor. The way in which non-magnetic scattering centers
affect the spin degrees of freedom is via the associated spin-orbit
interaction as described by the so-called Elliott-Yafet 
theory\cite{elliott,yafet}
Hence, if $v$ is the non-magnetic impurity potential,\cite{note1} the
corresponding spin-orbit scattering is proportional to $v_{\rm so}=v\delta g$,
where $\delta g$ is the shift of the $g$-factor. The actual value of
$\delta g$ depends on the wave function penetration
into the ions and the Fermi surface topology and it is a rather difficult 
problem.\cite{yafet} For copper-oxides the main contribution to $\delta g$ should
come from the $d$ orbital of Cu atoms for which $\delta g\simeq 0.1$.
Within the Born approximation, the spin-orbit scattering rate in the
normal state is therefore $1/\tau_{\rm so}\simeq
(\delta g)^2/\tau_{\rm imp}\ll 1/\tau_{\rm imp} $, where 
$1/\tau_{\rm imp}$ is the scattering rate due to $v$ alone. This is also
known as the Elliott-Yafet relation. However when the impurity
scattering is close to the unitary limit, as often advocated for impurity
doped high-$T_c$ superconductors, the Elliott-Yafet formula must be
generalized in order to include multi-scattering processes.
For a two-dimensional system with sufficiently
diluted impurity concentrations $n_i$, the solution of the normal state 
$T$-matrix equations for both $v$ and $v_{\rm so}$ leads to:\cite{grima1}
\begin{equation}
\label{ey}
\frac{1}{\tau_{\rm so}}=2\frac{1+c^2}{1+(2c/\delta g)^2}
\frac{1}{\tau_{\rm imp}},
\end{equation}
where $c=1/\pi N_0 v$, $1/\tau_{\rm imp}=2\Gamma/(1+c^2)$, 
$\Gamma=n_i/\pi N_0$ and
$N_0$ is the density of states per spin direction at the Fermi level.
In the weak scattering limit $c\gg 1$, Eq.(\ref{ey}) reproduces the
result of the Born approximation:
$1/\tau_{\rm so}=(\delta g)^2/2\tau_{\rm imp}\ll 1\tau_{\rm imp}$.
However for $c=0.1$, that is the value estimated in Ref.\onlinecite{hudson},
Eq.(\ref{ey}) leads to $1/\tau_{\rm so}\simeq 0.4/\tau_{\rm imp}$ when
$\delta g=0.1$, and in the extreme unitary limit:
$\lim_{c\rightarrow 0}1/\tau_{\rm so}= 2/\tau_{\rm imp}$ as long as 
$\delta g\neq 0$. Hence, when the impurity potential is strong, 
or more generally as long as $c \sim < \delta g$,
inevitably the spin-orbit interaction becomes as important as the
spin-independent coupling to the impurity. 

The above discussion suggests therefore that if non-magnetic impurities
in high-$T_c$ superconductors are close to the unitary limit, the
effect of spin-orbit coupling should be large. In particular, the
Zeeman response to an applied magnetic field should be deeply altered by 
the spin-mixing processes associated with $v_{\rm so}$ and eventually,
for sufficiently strong spin-orbit scattering, the Zeeman splitting 
should vanish. 
Here it is shown that for two-dimensional systems this 
conclusion is actually wrong: the Zeeman splitting resulting from the
solution of the $T$-matrix equation is much more robust
than that obtained within the Born approximation, and
at the unitary limit ($c/\delta g =0$) both $s$- and 
$d$-wave superconductors are fully Zeeman-splitted by an
applied in-plane magnetic field.

\section{Spin-orbit T-matrix}
\label{so}

For quasi-two dimensional systems the Zeeman response to an external 
magnetic field ${\bf H}$ should be best observed when ${\bf H}$ is
directed parallel to the conducting plane (for example the
Cu-O plane in copper-oxides), since in this case the coupling of ${\bf H}$
to the orbital motion of the electrons is minimized.\cite{makiparks}
Hence, the total hamiltonian is ${\cal H}={\cal H}_0+{\cal H}_{\rm imp}+
{\cal H}_{\rm so}$ where:
\begin{equation}
\label{h0}
{\cal H}_0=\sum_{{\bf k},\alpha}\epsilon({\bf k})
c^{\dagger}_{{\bf k}\alpha}
c_{{\bf k}\alpha}-h\sum_{{\bf k},\alpha}\alpha c^{\dagger}_{{\bf k}\alpha}
c_{{\bf k}\alpha} 
-\sum_{{\bf k}}\Delta({\bf k})(c^{\dagger}_{{\bf k}\uparrow}
c^{\dagger}_{-{\bf k}\downarrow}+c_{-{\bf k}\downarrow}c_{{\bf k}\uparrow}),
\end{equation}
where $\epsilon({\bf k})$ is the electron dispersion measured with respect
to the chemical potential, $\alpha$ is a spin index, 
$h=\mu_{\rm B}H$ and $\mu_{\rm B}$ is the 
Bohr magneton. In the following, it is assumed that the charge carriers are
confined to move in the $x$-$y$ plane so that ${\bf k}\equiv(k_x,k_y)$ and
that the spins are directed along
and opposite to the direction of the magnetic field, fixed to lie along
the $x$-direction: ${\bf H}=H\hat{x}$.
For $s$-wave superconductors $\Delta({\bf k})=\Delta$ while for $d$-wave
superconductors $\Delta({\bf k})=\Delta\cos(2\phi)$ where $\phi$ is the polar
angle in the $k_x-k_y$ plane. 
Without loss of generality, here $\Delta$
is used as an input parameter although it should be calculated self-consistently
from a suitable gap equation. Moreover, for simplicity, local variations of the order 
parameter are neglected. 
The impurity and spin-orbit hamiltonians, ${\cal H}_{\rm imp}$ and
${\cal H}_{\rm so}$, are given by:
\begin{equation}
\label{himp}
{\cal H}_{\rm imp}=v\sum_{{\bf k},{\bf k}',i}\sum_{\alpha}
e^{-i({\bf k}-{\bf k}')\cdot {\bf R}_i}
c^{\dagger}_{{\bf k}\alpha}c_{{\bf k}'\alpha},
\end{equation}
\begin{equation}
\label{hso}
{\cal H}_{\rm so}=i\frac{\delta g v}{k_F^2}\sum_{{\bf k},{\bf k}',i}
e^{-i({\bf k}-{\bf k}')\cdot {\bf R}_i}
([{\bf k}\times {\bf k}']\cdot \hat{{\bf z}})
(c^{\dagger}_{{\bf k}\uparrow}c_{{\bf k}'\downarrow}+
c^{\dagger}_{{\bf k}\downarrow}c_{{\bf k}'\uparrow}),
\end{equation}
where ${\bf R}_i$ denotes the random positions of the impurities
and $k_F$ is the Fermi momentum. Note that because of two-dimensionality,
the spin-orbit matrix element of Eq.(\ref{hso}) is proportional to
$\sigma_z$ and, since the spins are quantized along the $x$ direction,
the spin-orbit scattering is therefore always accompanied
by spin-flip processes. 
The Zeeman response for ${\cal H}_{\rm imp}=0$ and ${\cal H}_{\rm so}=0$
has already been considered in Ref.\onlinecite{yang1} for $d$-wave superconductors
and in Ref.\onlinecite{ghosh} for mixed symmetries of the order parameter. 
The inclusion of ${\cal H}_{\rm imp}$, which
however does not mix the spin states, has been studied in Ref.\onlinecite{won}.
The total hamiltonian ${\cal H}={\cal H}_0+{\cal H}_{\rm imp}+
{\cal H}_{\rm so}$ for $d$-wave symmetry has been considered 
in Ref.\onlinecite{grima2} within the
Born approximation for both impurity and spin-orbit scatterings.
Here, instead, the problem is generalized beyond the Born approximation
by solving the self-consistent 
$T$-matrix equation for both ${\cal H}_{\rm imp}$ and ${\cal H}_{\rm so}$.

The generalized Matsubara Green's function $G({\bf k},n)$ in the 
particle-hole spin space resulting from Eqs.(\ref{h0}-\ref{hso}) satisfies the
Dyson equation $G^{-1}({\bf k},n)=G_0^{-1}({\bf k},n)-\Sigma({\bf k},n)$,
where $G_0^{-1}({\bf k},n)=i\omega_n-\rho_3\epsilon({\bf k})-
\rho_2\tau_2\Delta({\bf k})-h\rho_3\tau_3$ is the propagator resulting from
${\cal H}_0$. The Pauli matrices $\rho_i$ and $\tau_i$ ($i=1,2,3$)
act on the particle-hole and spin subspaces, respectively. Whitin the
self-consistent $T$-matrix approach, the self energy is
$\Sigma({\bf k},n)=n_i T({\bf k},{\bf k},n)$ where the $T$-matrix
is the solution of the following equation:
\begin{equation}
\label{t1}
T({\bf k},{\bf k}',n)=u({\bf k},{\bf k}')+
\sum_{{\bf k}''}u({\bf k},{\bf k}'')G({\bf k}'',n)T({\bf k}'',{\bf k}',n),
\end{equation}
and $u({\bf k},{\bf k}')=\rho_3v+i \delta g v [\hat{{\bf k}}\times 
\hat{{\bf k}'}]_z \tau_1$. Because of the momentum dependence of the spin-orbit
part of $u({\bf k},{\bf k}')$, the $T$-matrix can be splitted into
the impurity and spin-orbit contributions for both $s$- and $d$-wave
symmetries of the order parameter. Hence, $T({\bf k},{\bf k}',n)=
T_{\rm imp}(n)+T_{\rm so}({\bf k},{\bf k}',n)$ where
$T_{\rm imp}(n)=\rho_3v+\rho_3v\sum_{{\bf k}}G({\bf k},n)T_{\rm imp}(n)$
is the usual impurity $T$-matrix and:
\begin{equation}
\label{tso}
T_{\rm so}({\bf k},{\bf k}',n)=i\delta g v [\hat{{\bf k}}
\times\hat{{\bf k}}']_z\tau_1
+i\delta g v\sum_{{\bf k}''}[\hat{{\bf k}}\times\hat{{\bf k}}'']_z
\tau_1 G({\bf k}'',n)T_{\rm so}({\bf k}'',{\bf k}',n), 
\end{equation}
is the spin-orbit $T$-matrix. The solution of Eq.(\ref{tso}) is 
of the form:\cite{grima1} 
\begin{equation}
\label{tso1}
T_{\rm so}({\bf k},{\bf k}',n)=
i\delta g v[\hat{{\bf k}}\times{\bf t}(\hat{{\bf k}}',n)]\tau_1,
\end{equation}
where 
\begin{equation}
\label{tso2}
{\bf t}(\hat{{\bf k}},n)=\hat{{\bf k}}+i\delta g v \sum_{{\bf k}'}
\hat{{\bf k}}'\tau_1G({\bf k}',n)[\hat{{\bf k}}'\times
{\bf t}(\hat{{\bf k}},n)]_z.
\end{equation}
The above equation can be easily solved in terms of the components
$t_x(\hat{{\bf k}},n)$ and $t_y(\hat{{\bf k}},n)$ of the vector operator 
${\bf t}(\hat{{\bf k}},n)$:
\begin{eqnarray}
\label{tx}
t_x(\hat{{\bf k}},n)&=&A_{xy}^{-1}(n)\left[\hat{k}_x+i\delta g v
\hat{k}_y\sum_{{\bf k}'}(\hat{k}_x)^2\tau_1 G({\bf k}',n)\right], \\
\label{ty}
t_y(\hat{{\bf k}},n)&=&A_{yx}^{-1}(n)\left[\hat{k}_y-i\delta g v
\hat{k}_x\sum_{{\bf k}'}(\hat{k}_y)^2\tau_1 G({\bf k}',n)\right], 
\end{eqnarray}
where $A_{xy}^{-1}(n)$ and $A_{yx}^{-1}(n)$ are the inverse of the
following $4\times 4$ matrices:
\begin{eqnarray}
\label{axy}
A_{xy}(n)=1-(\delta g v)^2 \left[\sum_{{\bf k}}(\hat{k}_x)^2\tau_1 G({\bf k},n)
\right]\left[\sum_{{\bf k}}(\hat{k}_y)^2\tau_1 G({\bf k},n)\right], \\
\label{ayx}
A_{yx}(n)=1-(\delta g v)^2 \left[\sum_{{\bf k}}(\hat{k}_y)^2\tau_1 G({\bf k},n)
\right]\left[\sum_{{\bf k}}(\hat{k}_x)^2\tau_1 G({\bf k},n)\right].
\end{eqnarray}
Finally, from Eq.(\ref{tso1}), $T_{\rm so}({\bf k},{\bf k},n)$ reduces to:
\begin{equation}
\label{tso3}
T_{\rm so}({\bf k},{\bf k},n)=i\delta g v \hat{k}_x\hat{k}_y
\left[A_{yx}^{-1}(n)-A_{xy}^{-1}(n)\right]\tau_1
+(\delta g v)^2\sum_{{\bf k}'}\left[A_{yx}^{-1}(n)(\hat{k}_x\hat{k}_y')^2+
A_{xy}^{-1}(n)(\hat{k}_y\hat{k}_x')^2\right]\tau_1 G({\bf k}',n)\tau_1.
\end{equation}
Further analysis of the $T$-matrix problem requires the explicit inclusion
of the symmetry of the order parameter. This is done in the next subsections where
Eq.(\ref{tso3}) is solved for both $s$- and $d$-wave symmetries. 

\subsection{$s$-wave symmetry}
\label{swave}
The usual procedure to evaluate self-consistently the electron propagator 
is to guess the form of $G({\bf k},n)$ which, after being substituted
into $T_{\rm imp}(n)$ and $T_{\rm so}({\bf k},{\bf k},n)$, generates only
combinations of $\rho_i$ and $\tau_j$ matrices already contained in $G({\bf k},n)$.
The direct substitution of $G_0({\bf k},n)$ into $T_{\rm imp}(n)$ and 
$T_{\rm so}({\bf k},{\bf k},n)$ is a practical way to guess the correct form
of $G({\bf k},n)$ via the Dyson equation. When this is done,
it is easy to realize that when the symmetry of the order parameter 
is $s$-wave, $\Delta ({\bf k})=\Delta$, the two matrices $A_{xy}$ and $A_{yx}$
defined in Eqs.(\ref{axy},\ref{ayx}) become equal. Hence, the form of
the electron propagator for an $s$-wave symmetry of the order parameter reduces to:
\begin{equation}
\label{g1}
G^{-1}({\bf k},n)=i[\tilde{\omega}-i\tilde{h}\rho_3\tau_3]-
\rho_3[\tilde{\epsilon}({\bf k})-i\tilde{\Lambda}\rho_3\tau_3] 
-\rho_2\tau_2[\tilde{\Delta}-i\tilde{\Gamma}\rho_3\tau_3],
\end{equation}
where the frequency dependence of the tilded quantities is implicit.
The tilded quantities are obtained by substituting Eq.(\ref{g1}) into the equations
for the impurity and spin-orbit $T$-matrices and requiring self-consistency
via the Dyson equation. In general, the solution is very complicated but a
considerable simplification arises if infinite electron band-width and
particle-hole symmetry of the normal state electron dispersion are assumed.
In this case in fact several integrals over ${\bf k}$ average to zero,\cite{hirsch2} 
leading to the following self-consistent equations:
\begin{eqnarray}
\label{w1}
i\tilde{\omega}_{\pm}&=&i(\omega_n \pm ih)+\frac{\Gamma}{1+c^2}g_{\pm} 
+2\Gamma\frac{(2c/\delta g)^2g_{\mp}+g_{\pm}}
{1+(2c/\delta g)^4+2(2c/\delta g)^2(f_+ f_- -g_+ g_-)}, \\
\label{d1}
\tilde{\Delta}_{\pm}&=&\Delta+\frac{\Gamma}{1+c^2}f_{\pm} 
+2\Gamma\frac{(2c/\delta g)^2f_{\mp}+f_{\pm}}
{1+(2c/\delta g)^4+2(2c/\delta g)^2(f_+ f_- -g_+ g_-)},
\end{eqnarray}
where $\tilde{\omega}_{\pm}=\tilde{\omega}\pm i\tilde{h}$,
$\tilde{\Delta}_{\pm}=\tilde{\Delta}\pm i\tilde{\Gamma}$ and
$g_{\pm}=i\tilde{\omega}_{\pm}/\sqrt{\tilde{\Delta}_{\pm}+
\tilde{\omega}_{\pm}^2}$, $f_{\pm}=\tilde{\Delta}_{\pm}/
\sqrt{\tilde{\Delta}_{\pm}+\tilde{\omega}_{\pm}^2}$. 
To display the spin-mixing effect of the spin-orbit interaction,
equations (\ref{w1},\ref{d1}) are more conveniently rewritten in
terms of
$u_{\pm}=\tilde{\omega}_{\pm}/\tilde{\Delta}_{\pm}$:
\begin{equation}
\label{u1}
u_{\pm}=\frac{\omega_n\pm ih}{\Delta}
+2\frac{\Gamma}{\Delta}\left(\frac{c}{\delta g}\right)^2 
\frac{\frac{\displaystyle u_{\mp}-u_{\pm}}
{\displaystyle (1+u_{\mp}^2)^{1/2}}}
{1+\left(\frac{2c}{\delta g}\right)^4\!+2
\left(\frac{2c}{\delta g}\right)^2\!
\frac{\displaystyle 1+u_{\pm}u_{\mp}}
{\displaystyle (1+u_{\pm}^2)^{1/2}(1+u_{\mp}^2)^{1/2}}}.
\end{equation}
Apart for the trivial limit $u_{\pm}=(\omega_n \pm ih)/\Delta$
which holds true in the absence of spin-orbit interaction ($\delta g=0$),
the two spin channels $u_+$ and $u_-$ are coupled together. 
Within the Born approximation, $c/\delta g \gg 1$, equation (\ref{u1})
reduces to the two-dimensional version of the $u_{\pm}$ formula
found in classic literature:\cite{makiparks,fulde}
\begin{equation}
\label{uborn}
u_{\pm}=\frac{\omega_n\pm ih}{\Delta}
+\frac{1}{2}\Gamma\left(\frac{\delta g}{c}\right)^2
\frac{u_{\pm}-u_{\mp}}{(1+u_{\mp}^2)^{1/2}}.
\end{equation}
The novel feature
displayed by the more general expression (\ref{u1}) is that,
as the unitary limit $c/\delta g=0$ is approached, $u_+$
and $u_-$ becomes decoupled and the full Zeeman splitting 
$u_+-u_-=2ih/\Delta$ is recovered. In such a limit therefore, 
the $s$-wave superconductor is fully Zeeman-splitted as if the spin-orbit
scattering would be spin conserving. The same conclusion can be obtained
by calculating the zero temperature spin susceptibility $\chi_{\rm s}$
as inferred by the linear response theory. In fact, by including
the spin-vertex function consistent with the $T$-matrix 
formulation it is possible to show that:
\begin{equation}
\label{chi1}
\frac{\chi_{\rm s}}{\chi_{\rm n}}=1-\frac{\pi T}{\Delta}
\sum_n\frac{1}{1+(\omega_n/\Delta)^2}\frac{1}
{[1+(\omega_n/\Delta)^2]^{1/2}+\rho_{\rm so}},
\end{equation}
where $\chi_{\rm n}=2\mu_{\rm B}^2 N_0$, $T$ is the temperature and $\rho_{\rm so}=
(\Gamma/\Delta)(\delta g/c)^2/[1+(\delta g/2c)^2]^2$. At zero temperature and
for $\rho_{\rm so}<1$, Eq.(\ref{chi1}) reduces to:
\begin{equation}
\label{chi2}
\frac{\chi_{\rm s}}{\chi_{\rm n}}=1-\frac{1}{\rho_{\rm so}}
\left[\frac{\pi}{2}-\frac{\arccos (\rho_{\rm so})}{\sqrt{1-\rho_{\rm so}^2}}\right].
\end{equation}
For $\delta g/c\ll 1$, $\rho_{\rm so}\simeq
(\Gamma/\Delta)(\delta g/c)^2$ and Eq.(\ref{chi2}) becomes equal to the 
two-dimensional version of the Abrikosov-Gorkov
expression based on the Born approximation.\cite{abrigorkov}
Instead, for $c/\delta g \ll 1$, $\rho_{\rm so}\simeq
(\Gamma/\Delta)(2c/\delta g)^2$ and $\chi_{\rm s}/\chi_{\rm n}\simeq 2\pi (\Gamma/\Delta)
(2c/\delta g)^2 \ll 1$ which vanishes when $c/\delta g =0$.

The absence of spin-mixing contributions at the unitary limit can be interpreted
as a consequence of the fact that ${\cal H}_0+{\cal H}_{\rm imp}$ commutes with $S_x$ while
${\cal H}_{\rm so}$ commutes with $S_z$. Therefore, for weak spin-orbit
scattering ($c/\delta g \gg 1$) $S_x$ is a rather good quantum number and
${\cal H}_{\rm so}$ induces weak spin-flip processes leading to
coupled $u_{\pm}$ equations. The spin-decoupling at the unitary limit $c/\delta g \ll 1$,
could be explained by arguing that for very strong spin-orbit interaction
$S_z$ rather than $S_x$ is a good quantum number.
The Cooper pairs are then formed by electrons with opposite spins in the $z$
direction and the spin rigidity of the superconducting condensate is efficient against
spin-flip transitions induced by the magnetic field ${\bf H}=H\hat{{\bf x}}$.
In the limiting case of infinitely strong spin-orbit interaction, therefore
$H$ can only induce polarization of the quasiparticle excitations.
Note that, in case the magnetic field is directed along the $z$ direction,
the total hamiltonian then commutes with $S_z$ and the Zeeman response
of a $s$-wave superconductor becomes independent of the spin-orbit 
interaction for whatever value of $c/\delta g$. Of course, for a three-dimensional system
the above reasoning does no longer apply because the spin-orbit interaction
does not commute with any component of ${\bf S}$.

\subsection{$d$-wave symmetry}
\label{dwave}

For the above considerations to be
valid it is required only two-dimensionality and a singlet superconducting
condensate. Therefore in principle also a two-dimensional $d$-wave 
superconductor 
should exhibit spin-channels decoupling as $c/\delta g \rightarrow 0$. 
This is indeed so even if there are qualitative differences with respect to
$s$-wave superconductors since for $d$-wave symmetry the
spin-orbit scattering becomes pair breaking. \cite{notapb}
Again assuming particle-hole symmetry and an infinite electron band-width, 
for $\Delta ({\bf k}) \equiv\Delta (\phi)=\Delta \cos(2\phi)$ the electron 
propagator is of the form:
\begin{equation}
\label{gd}
G^{-1}({\bf k},n)=i(\tilde{\omega}-i\tilde{h}\rho_3\tau_3)-
\rho_3[\tilde{\epsilon}({\bf k})-i\tilde{\Lambda}\rho_3\tau_3]-\rho_2\tau_2[\tilde{\Delta}(\phi) 
-i\tilde{\Gamma}(\phi)\rho_3\tau_3+i\tau_1\tilde{\Omega}(\phi)],
\end{equation}
where $\tilde{\Delta}(\phi)=\tilde{\Delta}\cos(2\phi)$, 
$\tilde{\Gamma}(\phi)=\tilde{\Gamma}\cos(2\phi)$, and 
$\tilde{\Omega}(\phi)=\tilde{\Omega}\sin(2\phi)$. 
The origin of $\tilde{\Omega}(\phi)$ (absent in the $s$-wave case) stems from 
the fact that, for $d$-wave
symmetry, the two matrices $A_{xy}$ and $A_{yx}$ in Eqs.(\ref{axy},\ref{ayx})
are no longer equal, so that the term proportional to 
$\hat{k}_x\hat{k}_y=\sin(2\phi)$ in $T_{\rm so}({\bf k},{\bf k},n)$, 
Eq.(\ref{tso3}), is nonzero.
As for the $s$-wave case the
self-consistent Dyson equation can be expressed in terms of $\tilde{\omega}_{\pm}$
and $\tilde{\Delta}_{\pm}$, but now there is an additional equation
for $\tilde{\Omega}$:
\begin{eqnarray}
\label{w2}
i\tilde{\omega}_{\pm}&=&i(\omega_n \pm ih)+
\frac{\Gamma}{c^2-(g^0_{\pm})^2}g^0_{\pm}
+2\Gamma\frac{(2c/\delta g)^2g_{\mp}+(f_{\mp}^2-g_{\mp}^2)g_{\pm}}
{[(2c/\delta g)^2-g_+g_- - f_+f_-]^2-(g_+f_-+g_-f_+)^2}, \\
\label{d2}
\tilde{\Delta}_{\pm}&=&\Delta 
+2\Gamma\frac{(2c/\delta g)^2f_{\mp}-(f_{\mp}^2-g_{\mp}^2)f_{\pm}}
{[(2c/\delta g)^2-g_+g_- - f_+f_-]^2-(g_+f_-+g_-f_+)^2}, \\
\label{omega}
\tilde{\Omega}&=&-2\Gamma\frac{(2c/\delta g)(g_+f_-+g_-f_+)}
{[(2c/\delta g)^2-g_+g_- - f_+f_-]^2-(g_+f_-+g_-f_+)^2},
\end{eqnarray}
\begin{eqnarray}
\label{gd1}
g_{\pm}&=&\frac{2}{\pi N_0}\sum_{{\bf k}}\frac{\sin(\phi)^2}{2\pi}
\frac{i\tilde{\omega}_{\pm}[\tilde{\omega}_{\mp}^2+E_{\mp}({\bf k})^2]\mp2
i(\tilde{\omega}_+-\tilde{\omega}_-)\tilde{\Omega}(\phi)^2}
{[\tilde{\omega}_+^2+E_+({\bf k})^2][\tilde{\omega}_-^2+E_-({\bf k})^2]
-\tilde{\Omega}(\phi)^2\{(\tilde{\omega}_+-\tilde{\omega}_-)^2+
[\tilde{\Delta}_+(\phi)-\tilde{\Delta}_-(\phi)]^2\}}, \\
\label{fd1}
f_{\pm}&=&\frac{2}{\pi N_0}\sum_{{\bf k}}\frac{\sin(\phi)^2}{2\pi}
\frac{\tilde{\Delta}_{\pm}(\phi)[\tilde{\omega}_{\mp}^2+E_{\mp}({\bf k})^2]\mp
[\tilde{\Delta}_+(\phi)-\tilde{\Delta}_-(\phi)]\tilde{\Omega}(\phi)^2}
{[\tilde{\omega}_+^2+E_+({\bf k})^2][\tilde{\omega}_-^2+E_-({\bf k})^2]
-\tilde{\Omega}(\phi)^2\{(\tilde{\omega}_+-\tilde{\omega}_-)^2+
[\tilde{\Delta}_+(\phi)-\tilde{\Delta}_-(\phi)]^2\}},
\end{eqnarray}
where $E_{\pm}({\bf k})^2=\epsilon({\bf k})^2+
\tilde{\Delta}_{\pm}(\phi)^2+\tilde{\Omega}(\phi)^2$ and $g^0_{\pm}$ 
can be obtained from Eq.(\ref{gd1}) by setting $\sin(\phi)^2\rightarrow 1/2$. 
The off-diagonal contribution $\tilde{\Omega}$
defined in Eq.(\ref{omega}) is responsible for spin-mixing terms appearing in
$g_{\pm}$, $f_{\pm}$, and $g^0_{\pm}$. However,
at the unitary limit $c/\delta g =0$, $\tilde{\Omega}$ vanishes and
the above self-consistent equations reduce to:
\begin{eqnarray}
\label{w3}
i\tilde{\omega}_{\pm}&=&i(\omega_n \pm ih)-\frac{\Gamma}{g^0_{\pm}}+
2\Gamma\frac{g_{\pm}}{f_{\pm}^2-g_{\pm}^2}, \\
\label{d3}
\tilde{\Delta}_{\pm}&=&\Delta-
2\Gamma\frac{f_{\pm}}{f_{\pm}^2-g_{\pm}^2},
\end{eqnarray}
where 
\begin{equation}
\label{g}
g_{\pm}=2\int\frac{d\phi}{2\pi}\frac{i\tilde{\omega}_{\pm}\sin(\phi)^2}
{[\tilde{\Delta}_{\pm}(\phi)^2+\tilde{\omega}_{\pm}^2]^{1/2}},
\end{equation}
\begin{equation}
\label{f}
f_{\pm}=2\int\frac{d\phi}{2\pi}\frac{\tilde{\Delta}_{\pm}(\phi)\sin(\phi)^2}
{[\tilde{\Delta}_{\pm}(\phi)^2+\tilde{\omega}_{\pm}^2]^{1/2}}.
\end{equation}
The two spin channels $+$ and $-$ in 
Eqs.(\ref{w3},\ref{d3}) are now completely decoupled in analogy therefore with
the $s$-wave case treated before. 
However now even at $T=0$ the spin susceptibility is expected to
remain non-zero (as long as $\Gamma\neq 0$). This is due to the fact that,
although there are not
spin-mixing processes at $c/\delta g =0$, the pair-breaking effect of 
both impurity and spin-orbit scatterings leads to a finite density
of states at the Fermi level.\cite{hotta} In this situation therefore 
the Zeeman splitted density of states is a more direct evidence for the
spin decoupling effect at $c/\delta g \rightarrow 0$. 
This is shown in Fig. 1 where the two spin channels density of states, $N_{\pm}(\omega)$,
are plotted for $\Gamma=0.1$, $\delta g=0.1$, $h=0.2$ and for different values of $c$.
$N_{\pm}(\omega)$ is calculated numerically from:
\begin{equation}
\label{dos}
\frac{N_{\pm}(\omega)}{N_0}=-{\rm sgn}(\omega){\rm Im}\!\left[g^0_{\pm}(\omega)\right],
\end{equation}
where $g^0_{\pm}(\omega)$ is the analytic continuation on the real axis 
($i\omega_n\rightarrow\omega+i\delta$) of Eq.(\ref{gd1}) 
(with $\sin(\phi)^2\rightarrow 1/2$).\cite{noteint}
In Fig. 1a, $N_{\pm}(\omega)$ is calculated from the solution of the general equations
(\ref{w2})-(\ref{omega}), while, for comparison, 
the result for the Born approximation applied to the spin-orbit part 
of Eqs.(\ref{w2}-\ref{omega}) is shown in Fig. 1b.\cite{noteborn} 
Up to $c=0.1$ ($\delta g/c=1$) the general $T$-matrix solution and the 
Born approximation agree quite well, while already for $c=0.05$ 
($\delta g/c=2$) the splitted coherence peaks at $\omega/\Delta\simeq
\pm(1\pm h)$ are still quite visible in Fig. 1a and completely suppressed in Fig. 1b.
Since at the unitary limit $c/\delta g=0$ the spins are completely decoupled,
Eqs.(\ref{w3},\ref{d3}), the two spin density of states are identical and
shifted by $\pm h$ one respect to the other. This is in contrast to the Born
solution, Fig. 1b, for which the strong spin-mixing terms lead to a flat density
of states. It should be stressed that before reaching the $\delta g/c \gg 1$ limit,
the Born approximation predicts that
$d$-wave superconductivity is already completely destroyed.\cite{notetc} 
Therefore the results for $c=0$ in Fig. 1b should be considered just as a 
mathematical limit to be compared with the solutions of the $T$-matrix approach
of Fig. 1a.

\section{conclusions}
\label{concl}
In summary, it has been shown that when non-magnetic impurity scattering is 
close to the unitary limit, the associated spin-orbit interaction is not
small provided $c \sim < \delta g$ and must be treated beyond the simple
Born approximation. Within the self-consistent $T$-matrix approach,  
it has been demonstrated that the Zeeman splitting of both $s$- and
$d$-wave two-dimensional superconductors is much more robust than
that obtained by the Born approximation.
At the unitary limit $c/\delta g =0$, for which the spin-orbit 
coupling is maximum, a two-dimensional $s$-wave superconductor 
does not show any spin-mixing processes and
the Zeeman response coincides with that of a pure
superconductor. Also for $d$-wave superconductors the spin-orbit coupling
becomes effectively spin-conserving at $c/\delta g =0$, but 
in addition it induces pair breaking effects which must be added to 
those caused by the scalar impurities.\cite{grima1}

Let us comment now on the possible limitations of the present theory.
The calculation method here used is a standard one based on
a $T$-matrix approximation for diluted impurities.\cite{hirsch,sunmaki,hirsch2} 
However when applied
to two-dimensional systems like the copper-oxides, the standard
procedure is complicated by the appearance of singularities in the
electron self-energy.\cite{nerse} Contrary to the results based on the $T$-matrix
solution and on self-consistent approaches to deal with the 
singularities,\cite{ziegler} non-perturbative methods suggest that the density of states
of a $d$-wave superconductor actually vanishes non analytically at 
the Fermi level.\cite{pepin} The low energy behavior appears to be heavily
modified by the level spacing of a localization volume which leads to
the opening of a pseudogap in the low-lying single electron excitations.\cite{senthil} 
It should be however noted that discepancies between different approaches
affect only the very low-energy excitations, while for energies not much smaller
than $\Delta$ the $T$-matrix approach is quite realiable (for finite but small
impurity concentrations). In this respect, the main result of Fig. 1a ({\it i. e.}
the persistence of the Zeeman splitting of the coherence peaks in 
the density of states even when $c/\delta g$ is zero) should not be an artificial
feature of the $T$-matrix approximation. This conclusion is also sustained by
the quite general physical explanation of the Zeeman-splitting persistence at
the unitary limit proposed in Sec.\ref{swave} and from an analysis of the
singee spin-orbit impurity problem not reported here. Note that, for these same reasons, 
some standard simplifications employed in the present calculations
(infinite band-width, particle-hole symmetric electron dispersion and 
absence of local suppressions of the order parameter) should not affect too seriously
the main result.

\newpage

\begin{figure}
\epsfxsize=30.5pc
\epsfbox{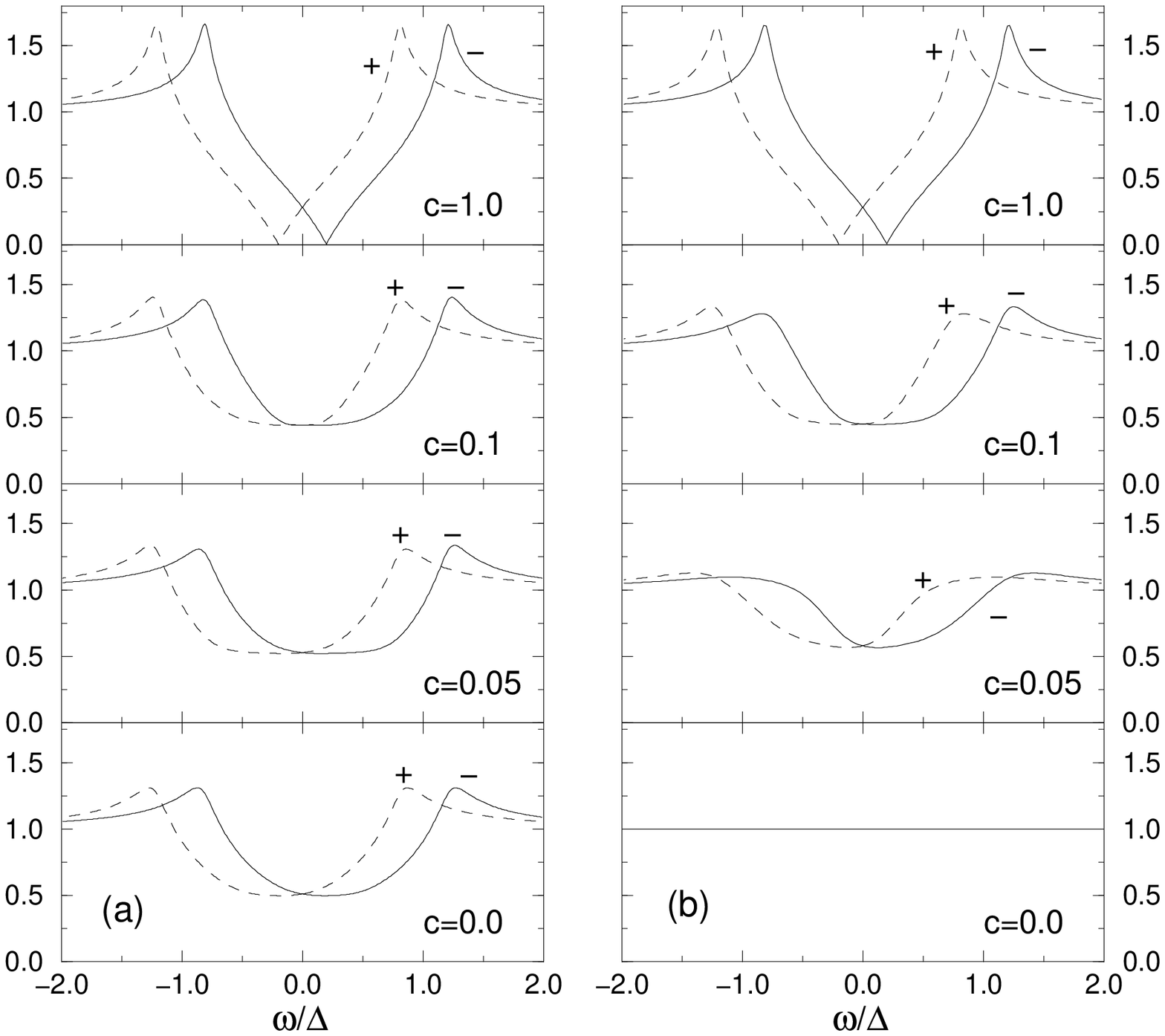}
%\centerline{\psfig{figure=fig4.eps,width=7.6cm}}
\caption{The Zeeman-split quasiparticle density of states $N_+(\omega)/N_0$
(dashed lines) and $N_-(\omega)/N_0$ (solid lines) for a $d$-wave superconductor
with $h=H/\Delta=0.2$, $\Gamma=0.1$, $\delta g=0.1$ and different values
of the scattering parameter $c$. (a): solution of the complete $T$-matrix
equations. (b): solution for the Born approximation to the spin-orbit coupling.}
\label{fig1}
\end{figure}


\begin{references}
\bibitem{note1}
For simplicity, the present analysis is restricted only to $s$-wave
impurity potentials of the form 
$V\sum_i \delta ({\bf r}-{\bf R}_i)$, where
${\bf R}_i$ are the random positions of the impurities.
$v$ is then given by $V/N_i$ where
$N_i$ is the number of scatterers.
\bibitem{notapb}
The pair-breaking effect of the spin-orbit scattering in $d$-wave
superconductors is obtained by solving self-consistently the gap equation.
This is done in Ref.\onlinecite{grima1} within the self-consistent
$T$-matrix approach without Zeeman magnetic fields ($h=0$) and in
Ref.\onlinecite{grima2} within the Born approximation for $h\neq 0$.
\bibitem{noteint}
Within the same approximation scheme employed for the $s$-wave case,
the integration over the energy in Eqs.(\ref{gd1},\ref{fd1}) can be easily done
analytically, but the final result is quite cumbersome and it is not reported here.
Hence, to obtain the quantities $g_{\pm}$, $f_{\pm}$, and $g^0_{\pm}$ it is
sufficient to perform numerically only an integration over the polar angle $\phi$.
\bibitem{noteborn}
For $\delta g/c\ll 1$, the last terms of Eq.(\ref{w2}) and Eq.(\ref{d2})
are replaced by $2\Gamma (\delta g/2c)^2g_{\mp}$, and 
$2\Gamma (\delta g/2c)^2f_{\mp}$, respectively.\cite{grima2}
Note that in the same approximation $\tilde{\Omega}\propto \Gamma(\delta g/2c)^3$
and can be neglected.
\bibitem{notetc}
From the solution of the gap equation within the
Born approximation of the spin-orbit coupling, it is found that
for $h=0$ the critical temperature $T_c$ goes to zero
as long as $(\delta g/2c)^2\ge \pi T_{c0}/2\gamma\Gamma-1/(1+c^2)$,
where $\gamma\simeq 1.781$ and $T_{c0}$ is the critical temperature
in the absence of impurities.\cite{grima1,grima2}
\end{references}
\end{document}